\documentclass[11pt]{combine}
\usepackage{amsmath,amsfonts,amssymb,amsthm,epsfig, graphicx, hyperref}
\usepackage[dvipsnames,table,dvipsnames*, svgnames*, hyperref]{xcolor}
\usepackage{natbib,amsmath,amssymb,amsthm,graphicx,setspace,paralist,booktabs,rotating,subcaption,float,color}
\usepackage{multirow}
\usepackage{fancyhdr}
\usepackage{bibunits}
\usepackage[small]{titlesec}
\newtheorem{theorem}{Theorem}
\newtheorem{corollary}{Corollary}
\newtheorem{proposition}{Proposition}

\newtheorem{lemma}{Lemma}
{
\theoremstyle{definition}
\newtheorem{definition}{Definition}
\newtheorem{example}{Example}

}
\newcommand{\beq}{\begin{equation}}
\newcommand{\eeq}{\end{equation}}
\newcommand{\beas}{\begin{align*}}
\newcommand{\eeas}{\end{align*}}
\newcommand{\bea}{\begin{align}}
\newcommand{\eea}{\end{align}}
\newcommand{\bei}{\begin{itemize}}
\newcommand{\eei}{\end{itemize}}
\newcommand{\ben}{\begin{enumerate}}
\newcommand{\een}{\end{enumerate}}
\newcommand{\bet}{\begin{theorem}}
\newcommand{\eet}{\end{theorem}}
\newcommand{\bel}{\begin{lemma}}
\newcommand{\eel}{\end{lemma}}
\newcommand{\bep}{\begin{proposition}}
\newcommand{\eep}{\end{proposition}}

\newcommand{\bed}{\begin{definition}}
\newcommand{\eed}{\end{definition}}
\newcommand{\bec}{\begin{corollary}}
\newcommand{\eec}{\end{corollary}}
\newcommand{\bex}{\begin{example}}
\newcommand{\eex}{\end{example}}

\numberwithin{equation}{section}

\newcommand{\T}[1]{{\text{T}}}
\usepackage{algorithm,algorithmic}

\begin{document}

%\begin{titlepage}

\title{\scshape Testing for publication bias in meta-analysis under Copas selection model}
\author{Rui Duan$^1$$^*$, Jin Piao$^2$$^*$, Arielle Marks-Anglin$^1$$^*$, Jiayi Tong$^1$,  Lifeng Lin$^3$,\\ Haitao Chu$^4$, Jing Ning$^5$ and Yong Chen$^1$ \\
$^1$Department of Biostatistics, Epidemiology and Informatics, \\
University of Pennsylvania,
Philadelphia, PA 19104\\
$^2$ Department of Preventive Medicine, University of Southern California, \\Los Angeles,  CA 90033\\
$^3$ Department of Statistics, Florida State University,  Tallahassee, FL 32306\\
$^4$ Division of Biostatistics, University of Minnesota,  Minneapolis, MN 55455\\
$^5$ Department of Biostatistics, The University of Texas MD Anderson Cancer Center,\\ Houston, TX 77030\\
$^*$ co-first authors
}
\date{}
%\author{Rong Ma, \ T. Tony Cai \ and \ Hongzhe Li}
%\date{}
\maketitle
\thispagestyle{empty}

	\begin{abstract}
In meta-analyses, publication bias is a well-known, important and challenging issue because the validity of the results from a meta-analysis is threatened if the sample of studies retrieved for review is biased. One popular method to deal with publication bias is the Copas selection model, which provides a flexible sensitivity analysis for correcting the estimates with considerable insight into the data suppression mechanism. However, rigorous testing procedures under the Copas selection model to detect bias are lacking. To fill this gap, we develop a score-based test for detecting publication bias under the Copas selection model. We reveal that the behavior of the standard score test statistic is irregular because the parameters of the Copas selection model disappear under the null hypothesis, leading to an identifiability problem. We propose a novel test statistic and derive its limiting distribution. A bootstrap procedure is provided to obtain the p-value of the test for practical applications. We conduct extensive Monte Carlo simulations to evaluate the performance of the proposed test and apply the method to several existing meta-analyses.
	\bigskip
	
\end{abstract}
%%%%%%%%%%%%%%%%%%%%%%%%%%%%%

\section{Introduction}\label{sec:intro}
The rapid growth of evidence-based medicine has led to substantially increased attention towards meta-analysis, which combines statistical evidence from multiple studies to improve power and precision \citep{cohn2003meta,jacksonmultivariate}. \textcolor{black}{A common and challenging issue in the use of meta-analysis is small-study effects (SSE), which undermines the validity of results from a standard meta-analysis \citep{egger1997bias, sterne2000publication,yin2019simulation}. SSE refer to the phenomenon that smaller studies show different, often larger, treatment effects than larger studies. Common reasons for SSE include publication bias, choice of outcome measure (eg. odds ratios or probabilities) and clinical heterogeneity. While other sources of SSE can be accounted for through subgroup analysis (in the case of clinical heterogeneity) and variance-stabilizing transformation of the outcome, publication bias is particularly concerning as it involves drawing conclusions on an incomplete, selective body of evidence. It is defined as the publication of studies depending on the direction and statistical significance of results, as well as other potential information such as language (selective inclusion of studies published in English) and availability (selective inclusion of studies easily accessible to the researcher) \citep{dickersin2005publication}. Thus combining only the identified published studies may lead to incorrect, commonly optimistic conclusions. In the last two decades, several studies have shown that trials with significant or ``positive" findings are more likely to be published than trials with non-significant results \citep{hopewell2009publication, johnson2017reproducibility}. While a great deal of effort has been devoted to developing statistical methods to detect and correct for SSE, less work has been done on a testing approach specifically for publication bias.}
%{The current methods that account for publication bias can be roughly classified into two categories: methods based on graphs, and methods based on parametric modeling of the selection mechanism. In the first category, probably the most popular graphical method is the funnel plot, which assesses the asymmetry of a scatter plot of treatment effects estimated from individual studies against their precisions. {Under the premise that asymmetry in a funnel plot can suggest potential publication bias \citep{egger1997bias, sterne2000publication, sterne2001systematic}, statistical tests for asymmetry have been developed in the literature \citep{begg1994operating, egger1997bias, macaskill2001comparison, sterne2001systematic,Moreno2009BMC}.} Duval and Tweedie \citep{duval2000trim, duval2000nonparametric} further developed the ``trim and fill" method for adjusting the missing studies in meta-analysis. Though widely used, these graph-based tests rely on the key assumption that asymmetry is a proxy for publication bias.} 

{\textcolor{black}{The current testing methods for SSE  are graph-based and rely on the principle of asymmetry, which is considered as a proxy for SSE.} When treatment effects from individual studies are mapped against their corresponding precisions, as in the widely used funnel plot, the presence of asymmetry can suggest SSE \citep{egger1997bias, sterne2000publication, sterne2001systematic}. Thus regression tests \citep{begg1994operating, egger1997bias, macaskill2001comparison, sterne2001systematic,Moreno2009BMC} and rank-based tests \citep{duval2000nonparametric} have been developed to more formally detect scatter plot asymmetry.} However, asymmetry \textcolor{black}{alone does not distinguish between publication bias and other causes of SSE, incluing} induced correlation between the effect size and standard error due to clinical or methodological differences between studies, or the choice of outcome measure used \citep{sterne2011recommendations}. Furthermore, when asymmetry is the result of publication bias, the tests have limited power when the sample size is small or only a moderate amount of bias is present \citep{sterne2000publication}.

{As an alternative to graph-based approaches,} researchers have proposed several selection models that impose parametric distributions to explicitly characterize the underlying mechanism for publication \citep{hedges1984estimation, hedges1992modeling, iyengar2009sensitivity}. Most notably, a selection model was suggested in a series of papers by Copas and his colleagues \citep{copas1997inference, copas1999works, copas2000meta, copas2001sensitivity} {which assumes that the selection probability depends on both the effect size estimate and its standard error. Unlike the graph-based methods, the Copas selection model has an advantage of explicitly relating the pattern of observed study estimates and precisions to a publication mechanism}\textcolor{black}{, unrelated to other possible forms of correlation between the estimates and standard errors}. {However, as acknowledged by \cite{copas2001sensitivity}, the data often contain little information on the parameters that characterize the publication mechanism. {Direct inference on the complete set of parameters is challenging since the likelihood function approaches a plateau around a wide range of parameter values \citep{carpenter2009copas}. Alternatively, statistical inference on the Copas selection model is primarily conducted through sensitivity analysis} which considers a range of possible values for some parameters \citep{copas2001sensitivity,carpenter2009copas}.}

In \cite{copas2000meta}, the authors proposed a simple testing approach based on a parametric modelling of the linear relationship of the effect sizes and the standard errors, which does not fully utilize the features of the Copas selection model. There is limited work on a hypothesis testing approach that is directly based on Copas' model. A primary reason might be that the standard testing procedures (e.g., Wald test, score test or likelihood ratio test) cannot be applied directly due to some non-regularity issues. Specifically, it will be shown later in this paper that under the null hypothesis where there is no selection bias, some of the parameters in Copas selection model are not identifiable and the corresponding Fisher information matrix is singular. The non-identifiability problem has been considered by many theoretical statisticians and is sometimes referred to as the Davies problem \citep{davies1977hypothesis, davies1987hypothesis}. The singular information matrix problem has been considered by \cite{rotnitzky2000likelihood} and is often encountered in mixture model settings \citep{qin2011hypothesis, ning2014class, HongJASA2017}. In general, hypothesis testing procedures with the above problems are non-standard and have to be considered with special attention.

In this paper, we propose a novel score test to circumvent the above two non-regularity problems. We observe that although the information matrix is singular under the null, a submatrix of the information matrix is still positive definite. By fixing the non-identifiable parameters at pre-specified values, a score test can still be constructed. To avoid the dependence of the inference on the choice of pre-specified values, we propose a score test by taking the maximum of score test statistics over a grid of possible pre-specified values. We also investigate how the range of the grid and number of grid points impact the performance of the test. The proposed score test is shown to converge weakly to the supreme of a Gaussian process. To enable the hypothesis testing procedure, we also propose a parametric bootstrap procedure, which can be easily implemented in practice.

The remainder of this paper is organized as follows. Section $2$ introduces the Copas selection model and describes the proposed score test, its asymptotic distribution and a parametric bootstrap procedure. In Section $3$ we present  results from simulation studies to investigate the finite sample performance of the proposed test, its sensitivity to the choice of grids, and compare it with the existing tests.  In Section $4$, we apply the proposed test and existing tests to two case studies of systematic reviews.  A  discussion is provided in Section $5$.

\section{Method}
\subsection{Copas selection model}
In this subsection we restate the Copas selection model proposed by \cite{copas1997inference}, and describe a slight modification. Let $Y_i$ denote the reported effect size in the $i$-th study. We assume $Y_i$ follows a normal distribution with mean $\mu_i$ and variance $\sigma_i^2$, where $\sigma_i^2$ is the true study variance which is often not observed, and we observe  only an estimator $s_i^2$ of the true variance.   Suppose the underlying effect size of the study is $\mu_i$ and it is assumed to follow a normal distribution with mean $\mu$, the population averaged effect size, and variance $\tau^2$. The parameter $\tau^2$ describes the between study heterogeneity, while  $\sigma_i^2$ denotes the within-study sampling variance. 

%% split this paragraph and make it clearer.
We model the observed outcome of study $i$ using the usual random effects model as
\begin{equation}\label{E1}
Y_i=\mu_i+\sigma_i\epsilon_i, \quad \epsilon_i \sim N(0,1), \quad \mu_i\sim N(\mu,\tau^2), \quad i=1,2,\dots,n.
\end{equation}
Model (\ref{E1}) can also be written as
\[
Y_i=\mu+\tau u_i+\sigma_i\epsilon_i,\quad u_i\sim N(0,1),\quad \epsilon_i\sim N(0,1).
\]
According to the Copas selection model, the study $i$ is published if and only if a postulated latent variable $Z_i>0$, where this $Z_i$ can be written as
\begin{equation}\label{E2}
Z_i=\gamma_0+\gamma_1/s_i+\delta_i,
\end{equation}
with $\gamma_0$ and $\gamma_1$ being unknown parameters which control the marginal probability of observing a study and $\delta_i$ is a random variable following a standard normal distribution. The parameter $\gamma_0$ controls the overall proportion of selection and $\gamma_1$ controls how the selection depends on the size of the study. The sign of $\gamma_1$ is expected to be positive, since larger studies are more likely to be accepted for publication. Publication bias is modeled by assuming a correlation $\rho$ between $\epsilon_i$ and $\delta_i$, such that
\begin{equation}
\left(\begin{array}{c}
\epsilon_i \\
\delta_i \end{array}\right)\sim N\left\{\left(\begin{array}{c}
0 \\
0 \end{array}\right),\left(   \begin{array}{cc}
1 & \rho \\
\rho & 1 \end{array} \right)\right\}\label{bivariate}
\end{equation}
The correlation $\rho$ controls how the probability of observing a study is influenced by the effect size of a study. 
From equation (\ref{E2}), we can calculate the marginal probability of the $i$-th study with a standard error $s_i$ being published as
\begin{equation*}
\text{Pr}(Z_i>0|s_i)=\text{Pr}(\delta_i>-\gamma_0-\gamma_1/s_i)=\Phi(\gamma_0+\gamma_1/s_i),
\end{equation*}
where $\Phi(\cdot)$ is the cumulative distribution function of the standard normal distribution.

To make inference on the population averaged effect size $\mu$, we can construct the log-likelihood of the observed studies,
\begin{align}
l =& \sum_{i=1}^n \left\{ \log {f}(y_i|Z_i>0,s_i) \right\}
=\sum_{i=1}^n \left[ \log \left\{ {{\textrm{Pr}}(Z_i>0|y_i,s_i))f(y_i)\over {\textrm{Pr}}(Z_i>0|s_i)} \right\} \right] \nonumber
\\\propto& \sum_{i=1}^n \left[-\frac{1}{2} \log(\tau^2+\sigma_i^2)-\frac{(y_i-\mu)^2}{2(\tau^2+\sigma_i^2)}-\log\{\Phi(\gamma_0+\gamma_1/s_i)\}+\log\{\Phi(v_i)\}  \right]\label{E3},
\end{align}
where
\begin{equation*}
v_i=\frac{\gamma_0+\gamma_1/s_i+\rho s_i(y_i-\mu)/(\tau^2+\sigma_i^2)}{\sqrt{1-\rho^2s_i^2/(\tau^2+\sigma_i^2)}}.
\end{equation*}

In the above likelihood, the unknown parameters $\sigma^2_i$ can be approximated using the observed variables to reduce the number of unknown parameters. In \cite{copas2000meta},  a function of the reported sampling variance $s_i^2$ is used to approximate $\sigma^2_i$, based on the derivation from the conditional probability. However, we propose to directly use the reported sampling variance $s_i^2$ to approximate the unknown within-study variance $\sigma_i^2$.  In the publication process, the study is first conducted and then selected. Regardless of whether the study is suppressed or observed,  the value of $s_i^2$  is obtained before the publication process, which should provide a fairly good estimation of the true variance $\sigma_i^2$. By replacing $\sigma_i^2$ with $s_i^2$, the unknown parameters in the likelihood function are reduced to  $\mu,\tau,\rho,\gamma_0$ and $\gamma_1$.

\subsection{Testing for publication bias}
When the correlation parameter $\rho=0$, it indicates that $Y_i$ and $Z_i$ are uncorrelated. In this case, the quantity $\Phi(v_i)$ in the likelihood function (\ref{E3}) reduces to the marginal probability $\Phi(\gamma_0+\gamma_1/s_i)$. The likelihood then reduces to the likelihood of the univariate random-effects model. Therefore, the univariate random-effects meta-analysis leads to correct inference for the overall effect $\mu$ and between-study variance $\tau^2$ when the correlation $\rho$ is zero. On the other hand, when $\rho \neq 0$, the standard meta-analysis leads to biased estimation of the parameters $\mu$ and $\tau^2$. Thus, under the model specified by equations (2.1) and (2.2), testing for publication bias is equivalent to testing for $H_0: \rho=0$. 

However, the standard results for asymptotic tests may not apply because of the following challenges:\\
(N1) Parameters $\gamma_0$ and $\gamma_1$ are not identifiable under the null hypothesis;\\
(N2) The Fisher information matrix is singular under the null hypothesis.\\
The asymptotic distribution of the likelihood ratio test under the above conditions is often not a $\chi^2$ distribution. Hypothesis testing procedures are non-standard,   and have to be considered on a case-by-case basis \citep{davies1977hypothesis, davies1987hypothesis,rotnitzky2000likelihood}. For example, the test of homogeneity in mixture models or mixture regression models, genetic admixture models, case-control studies with contaminated controls and testing for partial differential gene expression in microarray studies \citep{qin2011hypothesis, ning2014class, HongJASA2017}. Naive tests ignoring the above non-regularity conditions can produce misleading results. For example, the Wald test based on $\hat{\rho}/\hat{se}(\hat{\rho})$ is invalid because the negative Hessian matrix of the log-likelihood is singular under the null hypothesis.

Motivated by the method in \cite{Rathouz1999Biometrika}, when $\gamma_0$ and $\gamma_1$ are fixed at some known numbers, the submatrix of the Fisher information matrix is non-singular, and consequently the non-regular conditions (N1) and (N2) are avoided. To account for these facts, we propose a score test constructed as:
\begin{equation}
T=\sup_{\gamma_0,\gamma_1}\left[Z(\gamma_0,\gamma_1,\hat{\mu},\hat{\tau}) \right]^2\label{E4}
\end{equation}
where $(\hat{\mu},\hat{\tau}^2)$ is the constrained MLE of $ (\mu,\tau^2 )$ under $\rho=0$ for a given pair of values for $ (\gamma_0,\gamma_1)$, 
\begin{equation}
Z(\gamma_0,\gamma_1,\hat\mu,\hat\tau^2)=\frac{\sum_{i=1}^N S_i(\gamma_0,\gamma_1,\hat\mu,\hat\tau^2)}{\left[ \sum_{i=1}^N \hat I_{\rho|\mu,\tau}(\gamma_0,\gamma_1,
	\hat\mu,\hat\tau,0)\right]^{1/2}},\label{E5}
\end{equation}
\begin{equation*}
S_i(\gamma_0,\gamma_1,\mu,\tau)=\frac{\partial}{\partial \rho}l(\gamma_0,\gamma_1,\mu,\tau^2,\rho; y_i, s_i)|_{\rho=0},
\end{equation*}
and  $\hat I_{\rho|\mu,\tau}(\gamma_0,\gamma_1,
\mu,\tau,\rho)$ is the estimated partial information matrix where its explicit form is defined in the Supplementary Material.

%\begin{align*}
%&Z(\gamma_0,\gamma_1,\mu,\tau^2)=\frac{\sum_{i=1}^N S_i(\gamma_0,\gamma_1,\mu,\tau^2)}{\left[ \sum_{i=1}^N Var\{S_i(\gamma_0,\gamma_1,\mu,\tau^2)\}\right]^{1/2}},\\ 
%& \text{  and  }S_i(\gamma_0,\gamma_1,\mu,\tau)=\frac{\partial}{\partial \rho}\log L(\gamma_0,\gamma_1,\mu,\tau^2,\rho; y_i, s_i)|_{\rho=0}.
%\end{align*}
This test statistic is to calculate the score test statistic $Z(\gamma_0,\gamma_1,\hat{\mu},\hat{\tau})$ for a given pair of values for $\gamma_0$ and $\gamma_1$, and then construct $T$ by taking the supreme value of $Z(\gamma_0,\gamma_1,\hat{\mu},\hat{\tau})$ over a grid of $\gamma_0$ and $\gamma_1$. The range of $\gamma_0$ and $\gamma_1$ could be decided similarly to the sensitivity analysis in the Copas model. Usually, the range for $\gamma_0$ could be $[-2,2]$, and the range for $\gamma_1$ depends on the range of $s_i$ to make sure that the probability of the largest study being published is greater than a certain probability threshold (e.g. 0.9) and the probability of the smallest study being published is less than a certain probability threshold (e.g. 0.1). Additional discussion regarding how to choose the grid can be found in Appendix C of the Supplementary Materials. Deriving the asymptotic distribution of $T$ can be challenging and it may not be of a simple form for practical use. In such a case, a bootstrap procedure can be used if the asymptotic distribution is well behaved.

\subsection{Asymptotic results and a simple parametric bootstrap procedure}
In this subsection, we provide the asymptotic distribution of the proposed test, followed by a parametric bootstrap procedure for practical application.
\begin{theorem}\label{tm}
	Assume that the regularity conditions $(A_1)\sim(A_3)$ in Appendix A of the Supplementary Materials hold. Under the null hypothesis that $\rho = 0$, the test statistic $T$ converges weakly to $sup_{\gamma_0, \gamma_1}\{Z(\gamma_0, \gamma_1)^2\}$, where $Z(\gamma_0, \gamma_1)$ is a Gaussian process with variance 1 and the autocorrelation \\
	$$R (\gamma_0,\gamma_1;\gamma_0',\gamma_1' ) = \lim_{n\rightarrow\infty}E\left\{\frac{S_i(\gamma_0,\gamma_1,\mu,\tau,0)S_i(\gamma_0',\gamma_1',\mu,\tau,0)}{I_{\rho|\mu,\tau}(\gamma_0,\gamma_1,
		\mu,\tau,0)^{\frac{1}{2}}I_{\rho|\mu,\tau}(\gamma_0',\gamma_1',\mu,\tau,0)^{\frac{1}{2}}}\right\}, $$
	where $I_{\rho|\mu,\tau}$ is the conditional information defined in the Appendix B of the Supplementary Material. 
\end{theorem}
Theorem \ref{tm} provides the limiting distribution of the proposed test statistic $T$, and also provides the theoretical foundation for the following parametric bootstrap procedure which offers a simpler way to obtain a p-value for the test. 

\begin{enumerate}
	\item {For $n$ observed studies of a meta-analysis}, calculate the test statistic in equation~(\ref{E4}), denoted by $T$.
	\item Estimate $(\hat{\mu}_0,\hat{\tau}_0)$, which is the MLE of $(\mu, \tau)$ under the null hypothesis of $\rho=0$.
	\item Generate  $B_{boot}$ independent bootstrap samples (e.g. $B_{boot}$=200), each consisting of $n$ observations. Specifically, for $i \in \{1, \dots n\}$ we generate  $\epsilon_i \sim N(0,1)$ independently, and $s_{i,Boot}$ are drawn from the within-study standard errors of the original data with replacement. Then the bootstrap sample is constructed by sampling $\mu_i \sim N(\hat{\mu}_0,\hat{\tau}_0^2)$, and $y_{i,Boot}=\mu_i+s_{i,Boot} \epsilon_i.$
	\item Evaluate the score test statistic for each bootstrap sample. Denote $T^* = (T_1^*,\dots,T_{B_{boot}}^*)$.
	\item The p-value is computed by  $ \textup{p-value} = \frac{\#\{T^*>T\}}{B_{boot}}$.
	%\item If $p$ is smaller than the significant level (e.g. 0.05), then we reject the null hypothesis that there is no publication bias.
\end{enumerate}
In meta-analysis, the {number of studies} is often small or moderate. Therefore the computation time for the above parametric bootstrap is reasonable.

\section{Simulation studies}
We conducted simulation studies to examine the performance of our score test statistic. In our simulation, we generated the observed within-study variance $s_i$ from the fold normal distribution $|N(0.25, 2)|$, and generate $u_i$ from $N(0,1)$. The between study variance $\tau^2$ was set to $0.01$ and the overall effect size $\mu$ was 0.4. The variables $\epsilon_i$ and $\delta_i$ were generated together from a bivariate normal distribution with correlation $\rho$ and variance $1$. The selection parameters were set to be $(\gamma_0,\gamma_1)=(-1,0.65)$. If $Z_i<0$, the study was suppressed. We generated $y_i$ and $s_i$ as above until there were $n$ observations. We used grid partition on the $(\gamma_0,\gamma_1)$ square region of [-2,2] $\times$ [0,2] by 2,500 grid points with row unit of 0.08 and column unit of 0.04. In order to reduce the computational task, we did not use all the points in the grid region. Instead, we randomly chose $p$ points from the grid to calculate the test statistic. We used the parametric bootstrap described in Section 2.3 to obtain the p-value of the proposed test.

\subsection{Investigating the influence of number of points from grids on the statistical power of the proposed test}
\begin{figure}\label{points}
	\centering
	\caption{Empirical rejection rates (\%) of the proposed test $T$ for testing of publication bias under different sample size ($n$), correlation parameter ($\rho$) and number of grid points ($p$) at nominal levels $\alpha=0.05$ and $0.10$. 	For type I error setting (i.e., $\rho=0$), all rejection rates were calculated from 10,000 simulations. For power setting (i.e., $\rho\neq 0$), all rejection rates were calculated from 500 simulations. The last column on the right $(1^*)$ stands for rejection rates of the test statistic calculated from equation~(\ref{E5}) with $(\gamma_0, \gamma_1)$ fixed at the true value.}
	\includegraphics[scale=0.6]{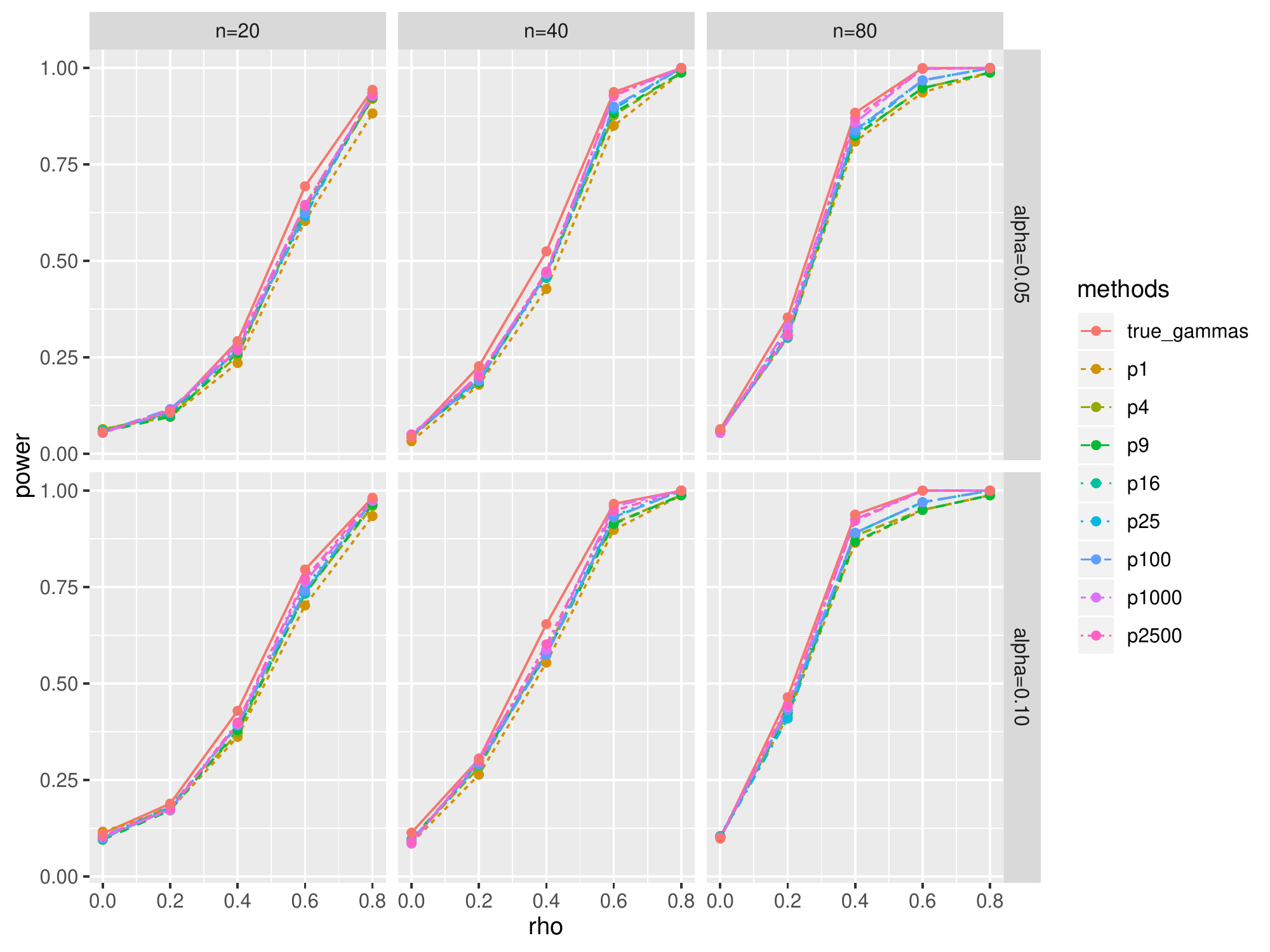}
\end{figure}

We first studied the influence of number of grid points $p$ on the performance of the test in terms of both type I error and power. Figure 1 presented the empirical rejection rates (i.e., type I error and power) under various numbers of grid points and sample size settings.  In order to evaluate the loss of power due to the lack of knowledge on the parameters $(\gamma_0, \gamma_1)$ that characterize the publication process, we included a test statistic calculated from equation~(\ref{E5}) with $(\gamma_0, \gamma_1)$ fixed at the true value. This test, although not practical in real application due to the unknown truth of $(\gamma_0, \gamma_1)$, can serve as a gold standard that informs the best possible statistical power under the Copas model. 

Figure 1 revealed the following interesting findings. First, under the null hypothesis of no publication bias, i.e. $\rho=0$, for all scenarios with different numbers of grid points, the empirical type I errors were close to the nominal level. This suggested that the parametric bootstrap described in Section 2.3 performed well. Secondly, by selecting more grid points, the statistical power of the proposed test increased. For example, when the sample size $n=40$ and $\rho=0.6$, the power increased from 85.1\% with $p=1$ to 93.6\% with $p=2500$. For most cases, the increase of power was rather limited when $p$ changed from $9$ to $2500$. Thirdly, by comparing the empirical power of the proposed test with the power of the test with $(\gamma_0, \gamma_1)$ fixed at the true value (i.e., the last column), we found that the proposed test only suffered a minimal loss of power due to the unknown $(\gamma_0, \gamma_1)$. These findings suggested that in practice, the proposed test can be performed by taking several random points from the grid (e.g. $n_g=10$), and it can still ensure a close statistical power to the ideal test with known $\gamma_0$ and $\gamma_1$.

\subsection{Comparing the proposed test with existing methods under the Copas model }

We compared the proposed test with the Trim and Fill test,  Egger's test, and the test proposed in \cite{copas2000meta}. From the results presented in Table 1, it was reasonable to consider the proposed test with only a few grid points. Thus we considered the proposed test with 9 points selected from the grid.

\begin{figure}\label{points2}
	\centering
	\caption{Comparison of statistical power (in \%) among proposed test with different number of grid points ($p$), Trim and fill test (TF) and Egger's test (Egger) at nominal levels $\alpha=0.05$ and $0.10$. All rejection rates were calculated from 500 simulations. }
	\includegraphics[scale=0.6]{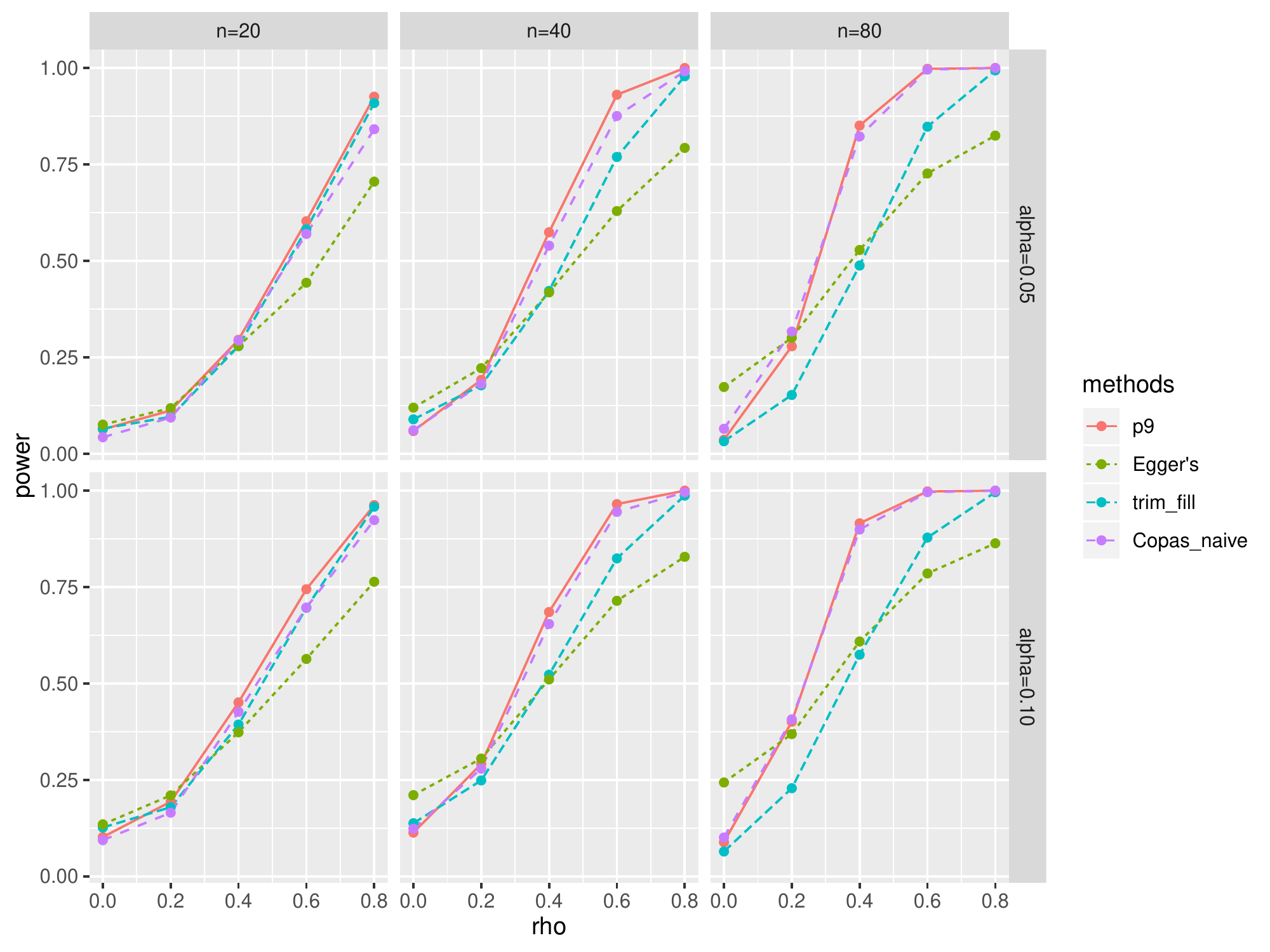}
\end{figure}

Figure 2 showed the comparison of type I error and power. For all scenarios, the type I errors were well controlled for the proposed test, but were slightly inflated for Egger's test. For the power comparison, the proposed test had the highest power among all scenarios. The Trim and Fill method outperformed the Egger's test and Copas naive test when sample size was small, while the Copas naive test outperformed the Trim and Fill method and  Egger's test for larger sample sizes.

\subsection{Power analysis when Copas model is not the true model}

In the Copas model, Equation~(\ref{E1}) is the random-effects assumption commonly used in the area of meta-analysis, which captures both the normality of estimates from each study, and the heterogeneity across studies. Equation~(\ref{E2}) is a working model describing the publication process, which captures several main features. First, there is stronger suppression on smaller studies compared to larger studies. They therefore set the latent variable to be positively correlated to the precision of the estimate, i.e., $1/s_i$, which can be considered as a proxy of the size of the study.  Second, if there is publication bias, studies with larger effect sizes ($Y_i$) has higher chance to be published given the precision ($s_i$). In the Copas model, this is captured by the correlation $\rho$, where a larger $\rho$ leads to stronger correlation between $Y_i$ and $Z_i$.

It is interesting to investigate the sensitivity of the proposed method, which is based on the Copas model, when the Copas model is not the correct model. To come up with some alternatives of the Copas selection model, we retain the random-effects model in Equation~(\ref{E1}), and alter the latent variable $Z_i$. We also maintain the two features where stronger suppression on smaller studies compared to larger studies, and studies with larger effect sizes ($Y_i$) have  higher chances to be published given the precision ($s_i$) in the existence of publication bias.  We considered the following two alternative models.

The first model  modifies the functional relationship between $Z_i$ and the term $1/s_i$, by changing equation (2) to 
\begin{equation}\label{EE3}
Z_i=\gamma_0+\gamma_1/s_i^2+\delta_i,
\end{equation}   
so that $Z_i$ is a linear function of $1/s_i^2$, instead of $1/s_i$. This is possible since the variance $s_i^2$ is roughly proportional to sample size of study $i$, and could be a better proxy of the size of the study.

Secondly, we change the way of characterizing publication bias. Instead of assuming a correlation $\rho$, we let $Z_i$ to be directly related to $y_i/s_i$ in the existence of publication bias, by assuming
\begin{equation}\label{EE4}
Z_i=\gamma_0+\gamma_1/s_i+c\rho y_i/s_i.
\end{equation}   
where $c$ is some positive scaling constant. This model links $Z_i$ with the $z$-score $y_i/s_i$, as the selection process could be depending on the $z$-score directly. Under Equation~(\ref{EE4}), it still holds that $\rho =0$ means no publication bias and studies with larger effect sizes ($Y_i$) have higher chances to be published given the precision ($s_i$) in the existence of publication bias.  In our simulation, we let $\rho$ vary from $0$ to $0.8$ to match the previous setting (in fact, in this alternative model $\rho$ does not need to be within $0$ and $1$), and set $c=0.5$ to obtain a proper range for the power curve.

To investigate the power and type I error of the proposed test under misspecified models, we simulated data from (\ref{EE3}) and (\ref{EE4}), and compared the performance of our method with existing methods. The results are presented in Figure 3. Under these two settings where the Copas model is the misspecified model, the power of the proposed test with $9$ points from the grid had similar power as the Trim and Fill method when sample size was small, and outperformed other methods with the increase of sample size. The type I error of the proposed method was still well-controlled around the nominal levels.

\begin{figure}\label{mis}
	\centering
	\caption{Upper panel: Comparison of statistical power (in \%) among proposed test with number of grid points to be $9$, Trim and Fill test (TF) and Egger's test (Egger) at nominal levels $\alpha=0.05$ and $0.10$ under model \ref{EE3}. Lower panel: Comparison of statistical power (in \%) among proposed test with number of grid points to be $9$, Trim and Fill test (TF) and Egger's test (Egger) at nominal levels $\alpha=0.05$ and $0.10$ under model \ref{EE4}.  All rejection rates were calculated from 500 simulations. }
	\includegraphics[scale=0.6]{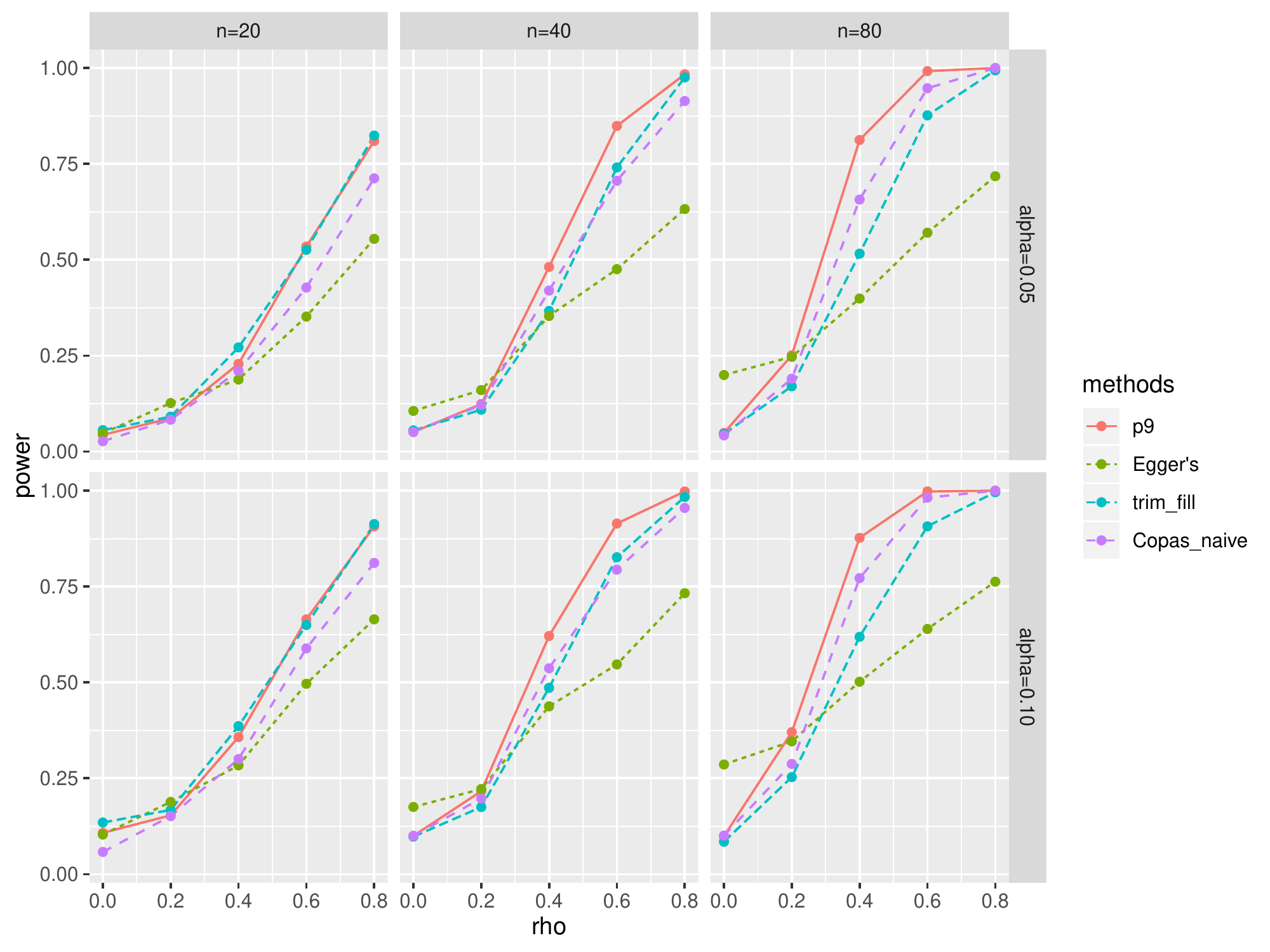}
	\includegraphics[scale=0.6]{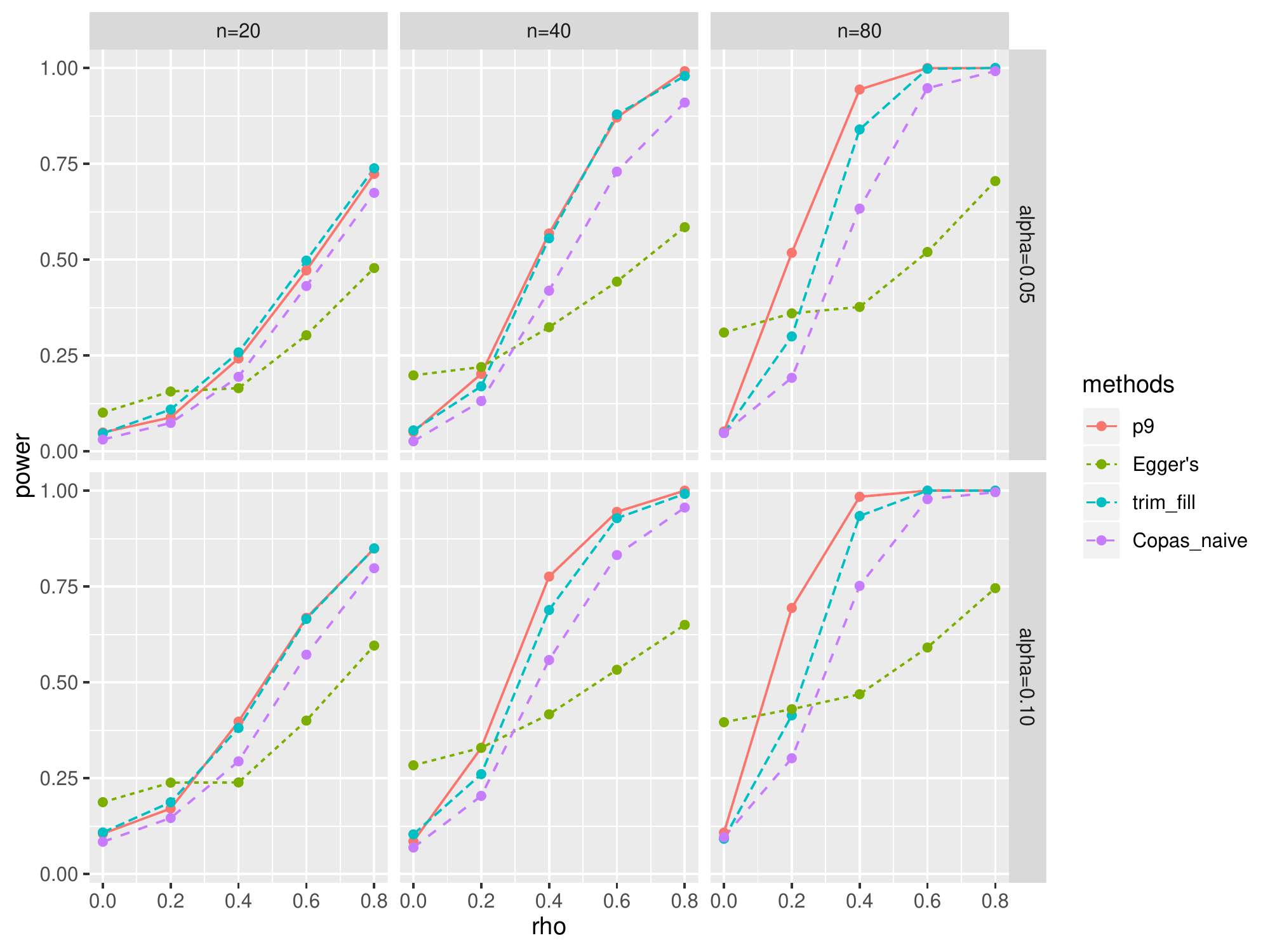}
\end{figure}

In summary, our simulation studies demonstrated that the proposed score test under Copas selection model behaves well with controlled Type I error and competitive statistical power relative to the existing tests. Furthermore, the implementation of the proposed test requires a small number of grid points and is computationally straightforward using parametric bootstrap.

\section{Data Analysis}
\subsection{A meta-analysis of  antidepressants comparing published studies with FDA reviews} Antidepressants are among the world’s most widely prescribed drugs, where many meta-analyses were conducted to synthesize evidence from existing trials to obtain a more reliable and generalizable results regarding treatment efficacy and safety \citep{barbui2011efficacy,cipriani2018comparative}.  By comparing trials of antidepressants that were published in the literature and unpublished trials obtained from the Food and Drug Administration (FDA), Turner et al. \cite{turner2008selective} identified strong evidence for selective bias due to publication towards results favoring active interventions. 

\begin{figure}
	\centering
	\caption{Left panel: ; Right panel}
	\includegraphics[scale=.45]{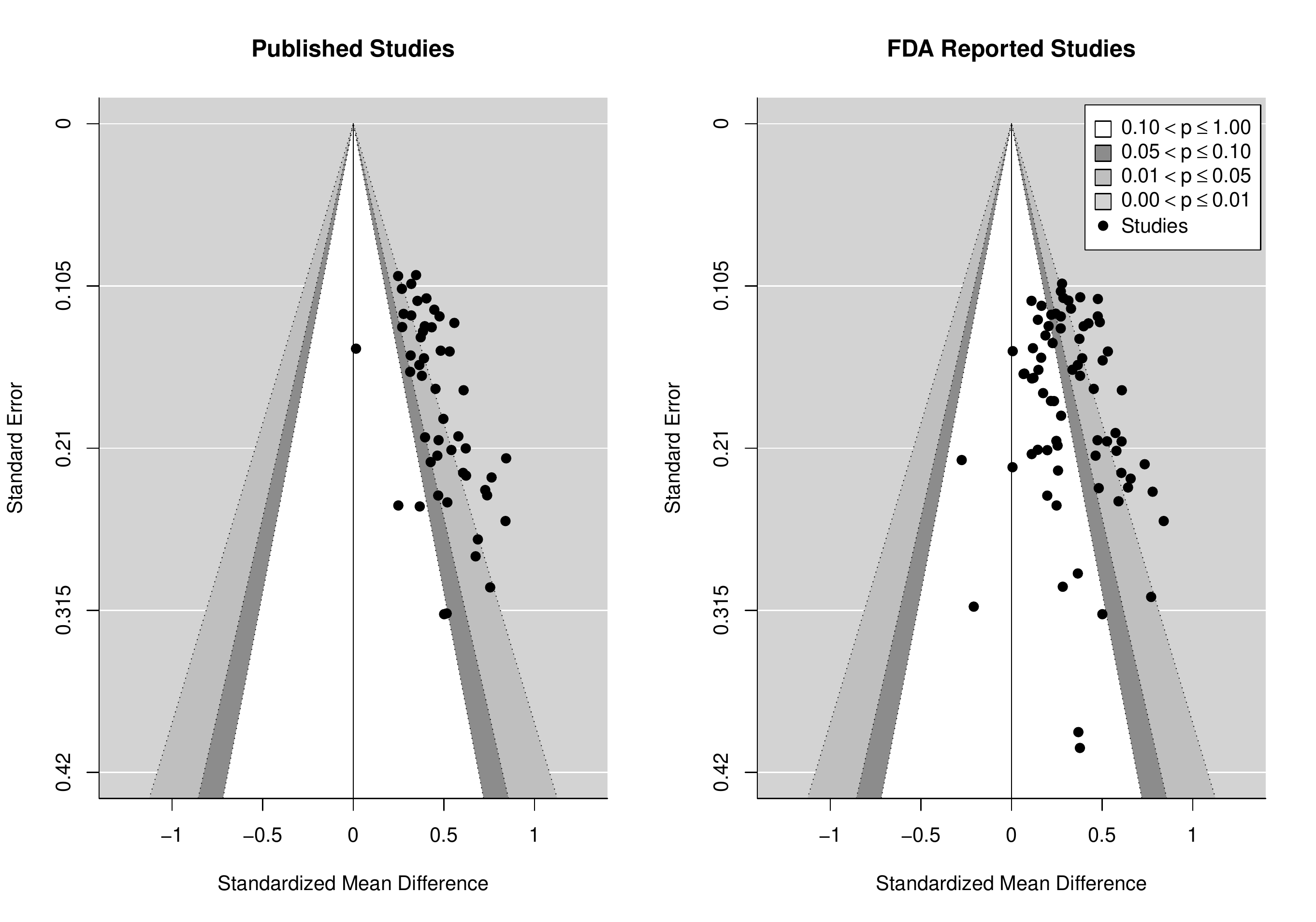}
	\label{turner}
\end{figure}

Although collecting information for unpublished trials is normally not feasible for meta-analyses, this unique dataset from \cite{turner2008selective} provide an opportunity to validate our method in an ideal situation, where the unpublished studies were actually observed. Among the 74 FDA-registered studies, 31\% were not published.  Figure \ref{turner} shows the contour enhanced funnel plots for all the published trials and all the registered trials.

By comparing the two funnel plots, we observed clear evidence that several non-significant results were suppressed for publication.  Applied on all the published studies, our test yielded a p-value  less than $0.01$, showing strong evidence of publication bias. Egger's test also captured the evidence for bias with a p-value  less than $0.01$. The Trim and Fill method, however, has a p-value of $0.25$, possibly driven by the left most trial with relatively large precision and small effect, which suggests its lack of power in the existence of outlying studies.

A more interesting investigation is to apply the above methods on all the FDA-registered trials. By including all the unpublished studies, our test had a p-value of $0. 24$ which suggested insufficient evidence for publication bias. Egger's test, however, had a p-value of $0.07$,  still showing marginal evidence for SSE which was likely not caused by publication bias. Trim and Fill test maintained a non-significant p-value of $0.50$, as the symmetry of the funnel plot is improved by adding all the unpublished results.

This case study demonstrated the capability our method for testing publication bias, which on the other than also showed the different focus of our test compared to existing methods. Based on the Copas model, our test is to detect the evidence for publication bias, where majority of the current methods, including Egger's test and the Trim and Fill method, is for testing SSE. Combing our test  with existing methods, we are able to provide a more comprehensive understanding of the potential sources of bias.

\subsection{ A meta-analysis on the effect of chewing gum after surgery to help recovery of the digestive system}

When patients have surgery on their abdomen, they are at risk to develop ileus, which is the inability of the intestine (bowel) to contract normally and move waste out of the body.  Chewing gum could be one possible way to prevent ileus,  since it tricks the body into thinking it is eating, and may help the digestive system to start working again. To investigate the effect of gum-chewing, Short et al. \cite{short2015chewing} did a systematic review, which included relevant trials focussed on people having bowel surgery, caesarean section, or other abdominal surgery types.  Using time to first flatus (TFF) as the primary outcome, slight improvement was identified comparing the gum-chewing groups to the control groups with an overall decrease of TFF ranging from 7.92 to 12.64 hours across different subgroups of surgery types. 

\begin{figure} 
	\centering
	\caption{Funnel plot of the meta-analysis on the effect of chewing gum after surgery to help recovery of the digestive system. In total 43 studies with 4224 participants were included. }
	\includegraphics[scale=.48]{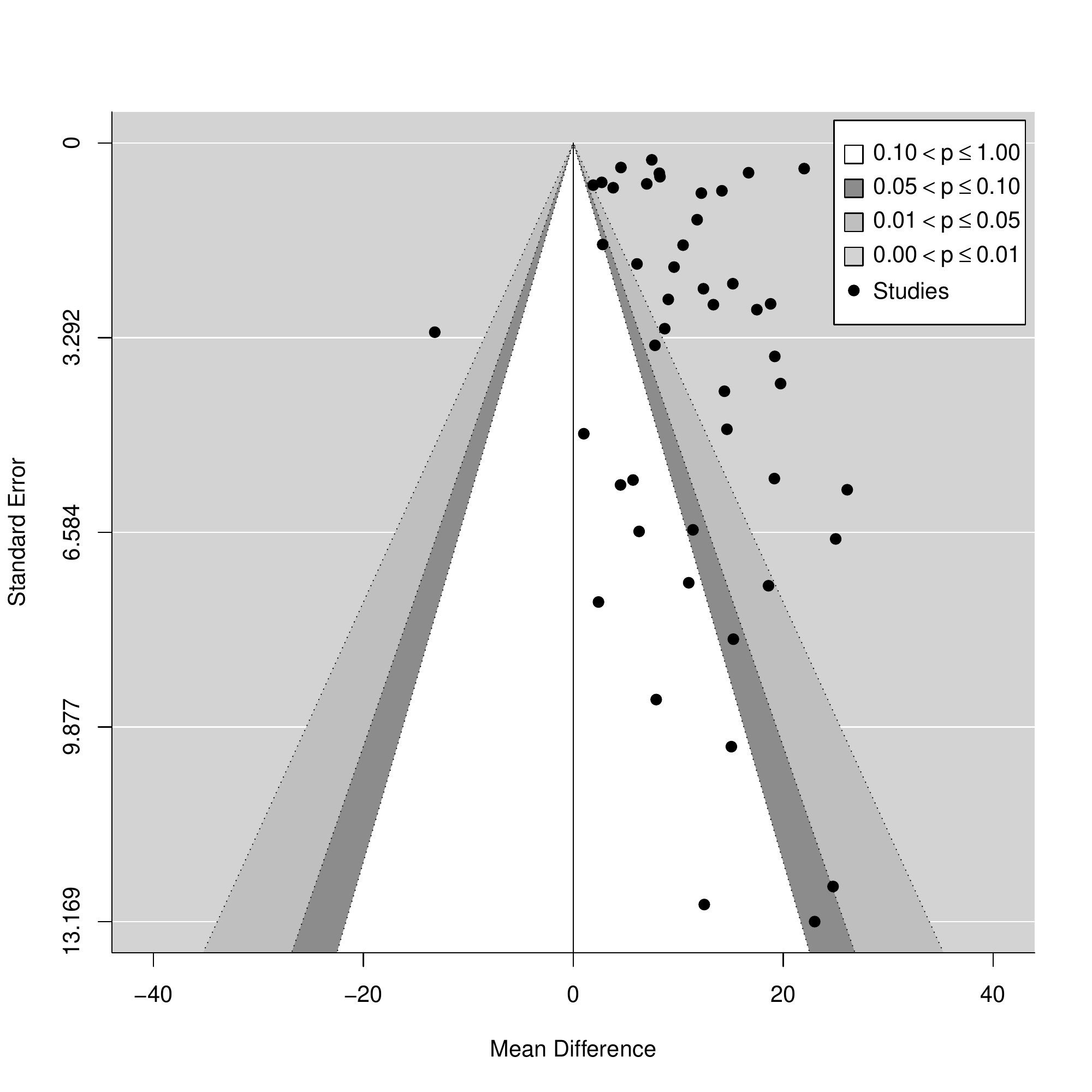}
\label{gum}
\end{figure}

To invesigate the potential risk for publication bias, we applied our method to the subgroup of trials for abdominal surgery types other than colorectal surgery and Caesarean, which was referred to the``Other Surgery" subgroup in \cite{short2015chewing}. This subgroup included in total 43 studies and  4224 participants. Figure \ref{gum} presented the funnel plot of the mean different of TFF comparing the control group to the gum-chewing group. From Figure \ref{gum}, we observed a possible trend of  selection by sign, where more studies with positive mean differences were observed. Many larger studies had mean difference near $0$, but only one study among the 43 studies had negative mean difference.

Applying our method with a $9$-point grid, we obtained a p-value of $0.03$, suggesting significance evidence for publication bias. The  Trim and Fill method had a p-value of $0.50$ and the Egger's test had a p-value of $0.12$. The non-significant p-values from the Trim-and-Fill method and the Egger's test were likely due to the lack of  funnel shape of the studies, where larger study were not observed to be more centered compared to the smaller studies, and the fact that the funnel plot was relatively symmetric.

In this study, our test was able to capture the evidence for potential publication bias, which was consistent with our observation from the funnel plot. We then conducted the Copas sensitivity analysis and obtained the adjusted overall mean difference to be $9.92 (7.82, 12.02)$, which was smaller compared to the overall mean difference $10.57 (8.47, 12.68)$ obtained by \cite{short2015chewing}. Although both results suggested chewing gum has significant effect in terms of decreasing the TFF, our testing approach revealed the risk for publication bias, where sensitivity analyses are recommended in order to explore the possible overall mean difference after adjusting for publication bias.   We believe that our method can be used as an complement of the existing methods for evaluating the risk of publication bias, which is crucial in assessing the reliability of conclusions from meta-analysis.

\section{Discussion}

Publication bias is a major threat to the validity of meta-analysis. In this paper, we proposed a testing procedure to detect publication bias under the Copas selection model, {as an alternative to current graph and symmetry based methods}. The standard score test under the Copas selection model is irregular since some model parameters disappear and the Fisher information matrix is singular under the null hypothesis. To circumvent these challenges, we constructed the test statistic by calculating the score test statistic for some fixed $\gamma_0$ and $\gamma_1$ on a grid, and taking the maximum value of score test statistics with different $\gamma_0$ and $\gamma_1$ over a grid. The asymptotic distribution of the test statistic was derived, and a parametric bootstrap procedure was developed to obtain p-values. We studied the empirical performance of our test statistic in terms of type I error and power under several scenarios in the simulation studies. Interestingly, we found that when increasing the number of grid points of $\gamma_0$ and $\gamma_1$, the statistical power increased and then became stable after some threshold. On the other hand, even when randomly choosing only a few points on the grid, the statistical power of our test was good relative to the best scenario in which the parameters in the Copas selection model $(\gamma_0, \gamma_1)$ were known. This empirical finding suggested some degree of optimality of the proposed test under the Copas selection model. 

There are several remarks on our proposed test. First, this proposed test is purely based on the Copas selection model, which is a parametric model of the publication mechanism.  {However, the real publication mechanism may be more complex and it is difficult to specify the true selection model based on the observed data \citep{copas2004bound}. Models such as \cite{gleser1996models} and \cite{rucker2010treatment} provide alternative ways for modeling selection bias. On the other hand, \cite{copas2001sensitivity} argued that despite the complexity of the true selection model, it is important to have a reasonably plausible model that can help to understand the publication mechanism. Empirical evaluation on a large number of meta-analyses has shown that the Copas model is capable of capturing the evidence of selection bias, as well as providing relatively clear interpretation \citep{carpenter2009empirical}. Our work filled an important research gap by proposing a hypothesis testing procedure for publication bias based on the well-accepted Copas selection model, and can be routinely applied to quantify the evidence of publication bias before conducting any sensitivity analysis using Copas model.}

We suggest the readers to use the proposed test as a way of evaluating the quality of the evidence from meta-analysis. Similar to how $I^2$ was designed for quantifying the heterogeneity for meta-analyses, our test statistic and the corresponding value can serve as a quantification of the potential risk of publication bias, based on the Copas selection model. Larger p-values correspond to lower risk of selective publishing. \textcolor{black}{If an extremely small p-value is observed, it alerts the investigators to a potential risk for bias. Ideally, this should prompt investigators to carefully search for unpublished and ongoing studies in international trial registers and drug approval agency websites, to obtain a more complete body of evidence for meta-analysis. However, if there are constraints on time and resources, or it is expected that many trials are unregistered (as with older studies), methods for bias reduction may also be used to account for publication bias. These include a Copas-based sensitivity analysis, the EM algorithm proposed by \cite{ning2017maximum}, and the non-parametric Trim and Fill method. However, as all post-hoc bias correction methods rely on untestable assumptions, we recommend treating them as sensitivity analysis rather than the main results of the meta-analysis.}

% (Need to add a few sentences on Schwarzer's work on implementing the Copas model and the practical challenges.)}  {\textcolor{red}{(need to refine this sentence as it is a key summary of our work.)}}
%Furthermore, the number of random points in the grid and the choice of the grid might influence the power of the test. However, as shown from the simulation study, the impact of the number of random points on statistical power is not substantial. In practice, it is suggested to choose a relatively wide range for the unknown parameter $\gamma_0$ and $\gamma_1$, and a reasonable number of points (e.g. 10-20). 

A second remark on our work is that we have focused on the case of univariate meta-analysis. In some applications, multiple outcomes are reported. For example, in diagnostic accuracy studies, both sensitivity and specificity of a diagnostic test are reported. In clinical trials comparing two treatments, outcomes related to treatment efficacy and safety are often reported. Models are developed for multivariate meta-analysis in order to jointly model the observed outcomes of interest. It is of interest to consider the publication bias problem when multiple outcomes are considered. Moreover, in multivariate meta-analyses, some studies only report part of the outcomes. This missing-data problem often causes bias in estimation if the missingness is related to the magnitude of the unobserved outcomes, which is known as outcome reporting bias \citep{kirkham2012multivariate}. By extending the Copas selection model into multivariate selection models, we can develop tests for detecting publication bias and outcome reporting bias in multivariate meta-analysis. In this case, building a flexible but parsimonious model for good model robustness and statistical power can be the key challenge in the modeling and construction of a testing procedure. This extension is currently under investigation and will be reported in the future.

\setstretch{1.24}
\bibliographystyle{chicago}
\bibliography{pb}

\begin{thebibliography}{}

\bibitem[\protect\citeauthoryear{Barbui, Cipriani, Patel, Ayuso-Mateos, and van
  Ommeren}{Barbui et~al.}{2011}]{barbui2011efficacy}
Barbui, C., A.~Cipriani, V.~Patel, J.~L. Ayuso-Mateos, and M.~van Ommeren
  (2011).
\newblock Efficacy of antidepressants and benzodiazepines in minor depression:
  systematic review and meta-analysis.
\newblock {\em The British Journal of Psychiatry\/}~{\em 198\/}(1), 11--16.

\bibitem[\protect\citeauthoryear{Begg and Mazumdar}{Begg and
  Mazumdar}{1994}]{begg1994operating}
Begg, C.~B. and M.~Mazumdar (1994).
\newblock Operating characteristics of a rank correlation test for publication
  bias.
\newblock {\em Biometrics\/}, 1088--1101.

\bibitem[\protect\citeauthoryear{Carpenter, R{\"u}cker, and
  Schwarzer}{Carpenter et~al.}{2009}]{carpenter2009copas}
Carpenter, J., G.~R{\"u}cker, and G.~Schwarzer (2009).
\newblock copas: An r package for fitting the copas selection model.
\newblock {\em The R Journal\/}~{\em 2}, 31--6.

\bibitem[\protect\citeauthoryear{Carpenter, Schwarzer, R{\"u}cker, and
  K{\"u}nstler}{Carpenter et~al.}{2009}]{carpenter2009empirical}
Carpenter, J.~R., G.~Schwarzer, G.~R{\"u}cker, and R.~K{\"u}nstler (2009).
\newblock Empirical evaluation showed that the copas selection model provided a
  useful summary in 80\% of meta-analyses.
\newblock {\em Journal of clinical epidemiology\/}~{\em 62\/}(6), 624--631.

\bibitem[\protect\citeauthoryear{Cipriani, Furukawa, Salanti, Chaimani,
  Atkinson, Ogawa, Leucht, Ruhe, Turner, Higgins, et~al.}{Cipriani
  et~al.}{2018}]{cipriani2018comparative}
Cipriani, A., T.~A. Furukawa, G.~Salanti, A.~Chaimani, L.~Z. Atkinson,
  Y.~Ogawa, S.~Leucht, H.~G. Ruhe, E.~H. Turner, J.~P. Higgins, et~al. (2018).
\newblock Comparative efficacy and acceptability of 21 antidepressant drugs for
  the acute treatment of adults with major depressive disorder: a systematic
  review and network meta-analysis.
\newblock {\em Focus\/}~{\em 16\/}(4), 420--429.

\bibitem[\protect\citeauthoryear{Cohn and Becker}{Cohn and
  Becker}{2003}]{cohn2003meta}
Cohn, L.~D. and B.~J. Becker (2003).
\newblock How meta-analysis increases statistical power.
\newblock {\em Psychological methods\/}~{\em 8\/}(3), 243.

\bibitem[\protect\citeauthoryear{Copas}{Copas}{1999}]{copas1999works}
Copas, J. (1999).
\newblock What works?: selectivity models and meta-analysis.
\newblock {\em Journal of the Royal Statistical Society: Series A (Statistics
  in Society)\/}~{\em 162\/}(1), 95--109.

\bibitem[\protect\citeauthoryear{Copas and Jackson}{Copas and
  Jackson}{2004}]{copas2004bound}
Copas, J. and D.~Jackson (2004).
\newblock A bound for publication bias based on the fraction of unpublished
  studies.
\newblock {\em Biometrics\/}~{\em 60\/}(1), 146--153.

\bibitem[\protect\citeauthoryear{Copas and Shi}{Copas and
  Shi}{2001}]{copas2001sensitivity}
Copas, J. and J.~Shi (2001).
\newblock A sensitivity analysis for publication bias in systematic reviews.
\newblock {\em Statistical Methods in Medical Research\/}~{\em 10\/}(4),
  251--265.

\bibitem[\protect\citeauthoryear{Copas and Shi}{Copas and
  Shi}{2000}]{copas2000meta}
Copas, J. and J.~Q. Shi (2000).
\newblock Meta-analysis, funnel plots and sensitivity analysis.
\newblock {\em Biostatistics\/}~{\em 1\/}(3), 247--262.

\bibitem[\protect\citeauthoryear{Copas and Li}{Copas and
  Li}{1997}]{copas1997inference}
Copas, J.~B. and H.~Li (1997).
\newblock Inference for non-random samples.
\newblock {\em Journal of the Royal Statistical Society: Series B (Statistical
  Methodology)\/}~{\em 59\/}(1), 55--95.

\bibitem[\protect\citeauthoryear{Davies}{Davies}{1977}]{davies1977hypothesis}
Davies, R.~B. (1977).
\newblock Hypothesis testing when a nuisance parameter is present only under
  the alternative.
\newblock {\em Biometrika\/}~{\em 64\/}(2), 247--254.

\bibitem[\protect\citeauthoryear{Davies}{Davies}{1987}]{davies1987hypothesis}
Davies, R.~B. (1987).
\newblock Hypothesis testing when a nuisance parameter is present only under
  the alternative.
\newblock {\em Biometrika\/}~{\em 74\/}(1), 33--43.

\bibitem[\protect\citeauthoryear{Dickersin}{Dickersin}{2005}]{dickersin2005publication}
Dickersin, K. (2005).
\newblock Publication bias: Recognizing the problem, understanding its origins
  and scope, and preventing harm.
\newblock {\em Publication Bias in Meta-Analysis\/}, 9--33.

\bibitem[\protect\citeauthoryear{Duval and Tweedie}{Duval and
  Tweedie}{2000}]{duval2000nonparametric}
Duval, S. and R.~Tweedie (2000).
\newblock A nonparametric trim and fill method of accounting for publication
  bias in meta-analysis.
\newblock {\em Journal of the American Statistical Association\/}~{\em
  95\/}(449), 89--98.

\bibitem[\protect\citeauthoryear{Egger, Smith, Schneider, and Minder}{Egger
  et~al.}{1997}]{egger1997bias}
Egger, M., G.~D. Smith, M.~Schneider, and C.~Minder (1997).
\newblock Bias in meta-analysis detected by a simple, graphical test.
\newblock {\em BMJ\/}~{\em 315\/}(7109), 629--634.

\bibitem[\protect\citeauthoryear{Gleser and Olkin}{Gleser and
  Olkin}{1996}]{gleser1996models}
Gleser, L.~J. and I.~Olkin (1996).
\newblock Models for estimating the number of unpublished studies.
\newblock {\em Statistics in medicine\/}~{\em 15\/}(23), 2493--2507.

\bibitem[\protect\citeauthoryear{Hedges}{Hedges}{1984}]{hedges1984estimation}
Hedges, L.~V. (1984).
\newblock Estimation of effect size under nonrandom sampling: The effects of
  censoring studies yielding statistically insignificant mean differences.
\newblock {\em Journal of Educational and Behavioral Statistics\/}~{\em
  9\/}(1), 61--85.

\bibitem[\protect\citeauthoryear{Hedges}{Hedges}{1992}]{hedges1992modeling}
Hedges, L.~V. (1992).
\newblock Modeling publication selection effects in meta-analysis.
\newblock {\em Statistical Science\/}, 246--255.

\bibitem[\protect\citeauthoryear{Hong, Chen, Ning, Wang, Wu, and Carroll}{Hong
  et~al.}{2017}]{HongJASA2017}
Hong, C., Y.~Chen, Y.~Ning, S.~Wang, H.~Wu, and R.~Carroll (2017).
\newblock Plemt: A novel pseudolikelihood based $\text{EM}$ test for
  homogeneity in generalized exponential tilt mixture models.
\newblock {\em Journal of the American Statistical Association\/}.

\bibitem[\protect\citeauthoryear{Hopewell, Loudon, Clarke, Oxman, and
  Dickersin}{Hopewell et~al.}{2009}]{hopewell2009publication}
Hopewell, S., K.~Loudon, M.~J. Clarke, A.~D. Oxman, and K.~Dickersin (2009).
\newblock Publication bias in clinical trials due to statistical significance
  or direction of trial results.
\newblock {\em The Cochrane Library\/}.

\bibitem[\protect\citeauthoryear{Iyengar and Greenhouse}{Iyengar and
  Greenhouse}{2009}]{iyengar2009sensitivity}
Iyengar, S. and J.~B. Greenhouse (2009).
\newblock Sensitivity analysis and diagnostics.
\newblock {\em The handbook of research synthesis and meta-analysis\/}, 417.

\bibitem[\protect\citeauthoryear{Jackson, Riley, and White}{Jackson
  et~al.}{2011}]{jacksonmultivariate}
Jackson, D., R.~Riley, and I.~White (2011).
\newblock Multivariate meta-analysis: Potential and promise.
\newblock {\em Statistics in Medicine\/}~{\em 30\/}(20), 2481--2498.

\bibitem[\protect\citeauthoryear{Johnson, Payne, Wang, Asher, and
  Mandal}{Johnson et~al.}{2017}]{johnson2017reproducibility}
Johnson, V.~E., R.~D. Payne, T.~Wang, A.~Asher, and S.~Mandal (2017).
\newblock On the reproducibility of psychological science.
\newblock {\em Journal of the American Statistical Association\/}~{\em
  112\/}(517), 1--10.

\bibitem[\protect\citeauthoryear{Kirkham, Riley, and Williamson}{Kirkham
  et~al.}{2012}]{kirkham2012multivariate}
Kirkham, J., R.~Riley, and P.~Williamson (2012).
\newblock A multivariate meta-analysis approach for reducing the impact of
  outcome reporting bias in systematic reviews.
\newblock {\em Statistics in Medicine\/}~{\em 31\/}(20), 2179--2195.

\bibitem[\protect\citeauthoryear{Macaskill, Walter, and Irwig}{Macaskill
  et~al.}{2001}]{macaskill2001comparison}
Macaskill, P., S.~D. Walter, and L.~Irwig (2001).
\newblock A comparison of methods to detect publication bias in meta-analysis.
\newblock {\em Statistics in medicine\/}~{\em 20\/}(4), 641--654.

\bibitem[\protect\citeauthoryear{Moreno, Sutton, Ades, Stanley, Peters, and
  Cooper}{Moreno et~al.}{2009}]{Moreno2009BMC}
Moreno, S., A.~Sutton, A.~Ades, T.~D. Stanley, J.~Peters, and N.~J. Cooper
  (2009).
\newblock Glutathione s-transferase (gst) m1, t1, p1, n-acetyltransferase (nat)
  1 and 2 genetic polymorphisms and susceptibility to colorectal cancer.
\newblock {\em BMC medical research methodology\/}~{\em 9\/}(1), 1.

\bibitem[\protect\citeauthoryear{Ning, Chen, and Piao}{Ning
  et~al.}{2017}]{ning2017maximum}
Ning, J., Y.~Chen, and J.~Piao (2017).
\newblock Maximum likelihood estimation and em algorithm of copas-like
  selection model for publication bias correction.
\newblock {\em Biostatistics\/}~{\em 18\/}(3), 495--504.

\bibitem[\protect\citeauthoryear{Ning and Chen}{Ning and
  Chen}{2014}]{ning2014class}
Ning, Y. and Y.~Chen (2014).
\newblock A class of pseudolikelihood ratio tests for homogeneity in
  exponential tilt mixture models.
\newblock {\em Scandinavian Journal of Statistics\/}.

\bibitem[\protect\citeauthoryear{Qin and Liang}{Qin and
  Liang}{2011}]{qin2011hypothesis}
Qin, J. and K.-Y. Liang (2011).
\newblock Hypothesis testing in a mixture case--control model.
\newblock {\em Biometrics\/}~{\em 67\/}(1), 182--193.

\bibitem[\protect\citeauthoryear{Rathouz and Liang}{Rathouz and
  Liang}{1999}]{Rathouz1999Biometrika}
Rathouz, P. and K.-Y. Liang (1999).
\newblock Reducing sensitivity to nuisance parameters in semiparametric models:
  a quasi-score method.
\newblock {\em Biometrika\/}~{\em 86\/}(4), 857--869.

\bibitem[\protect\citeauthoryear{Rotnitzky, Cox, Bottai, and Robins}{Rotnitzky
  et~al.}{2000}]{rotnitzky2000likelihood}
Rotnitzky, A., D.~R. Cox, M.~Bottai, and J.~Robins (2000).
\newblock Likelihood-based inference with singular information matrix.
\newblock {\em Bernoulli\/}~{\em 6\/}(2), 243--284.

\bibitem[\protect\citeauthoryear{R{\"u}cker, Schwarzer, Carpenter, Binder, and
  Schumacher}{R{\"u}cker et~al.}{2010}]{rucker2010treatment}
R{\"u}cker, G., G.~Schwarzer, J.~R. Carpenter, H.~Binder, and M.~Schumacher
  (2010).
\newblock Treatment-effect estimates adjusted for small-study effects via a
  limit meta-analysis.
\newblock {\em Biostatistics\/}~{\em 12\/}(1), 122--142.

\bibitem[\protect\citeauthoryear{Short, Herbert, Perry, Atkinson, Ness,
  Penfold, Thomas, Andersen, and Lewis}{Short et~al.}{2015}]{short2015chewing}
Short, V., G.~Herbert, R.~Perry, C.~Atkinson, A.~R. Ness, C.~Penfold,
  S.~Thomas, H.~K. Andersen, and S.~J. Lewis (2015).
\newblock Chewing gum for postoperative recovery of gastrointestinal function.
\newblock {\em Cochrane Database of Systematic Reviews\/}~(2).

\bibitem[\protect\citeauthoryear{Sterne, Egger, and Smith}{Sterne
  et~al.}{2001}]{sterne2001systematic}
Sterne, J.~A., M.~Egger, and G.~D. Smith (2001).
\newblock Systematic reviews in health care: Investigating and dealing with
  publication and other biases in meta-analysis.
\newblock {\em BMJ\/}~{\em 323\/}(7304), 101.

\bibitem[\protect\citeauthoryear{Sterne, Gavaghan, and Egger}{Sterne
  et~al.}{2000}]{sterne2000publication}
Sterne, J.~A., D.~Gavaghan, and M.~Egger (2000).
\newblock Publication and related bias in meta-analysis: power of statistical
  tests and prevalence in the literature.
\newblock {\em Journal of Clinical Epidemiology\/}~{\em 53\/}(11), 1119--1129.

\bibitem[\protect\citeauthoryear{Sterne, Sutton, Ioannidis, Terrin, Jones, Lau,
  Carpenter, R{\"u}cker, Harbord, Schmid, et~al.}{Sterne
  et~al.}{2011}]{sterne2011recommendations}
Sterne, J.~A., A.~J. Sutton, J.~P. Ioannidis, N.~Terrin, D.~R. Jones, J.~Lau,
  J.~Carpenter, G.~R{\"u}cker, R.~M. Harbord, C.~H. Schmid, et~al. (2011).
\newblock Recommendations for examining and interpreting funnel plot asymmetry
  in meta-analyses of randomised controlled trials.
\newblock {\em BMJ\/}~{\em 343}, d4002.

\bibitem[\protect\citeauthoryear{Turner, Matthews, Linardatos, Tell, and
  Rosenthal}{Turner et~al.}{2008}]{turner2008selective}
Turner, E.~H., A.~M. Matthews, E.~Linardatos, R.~A. Tell, and R.~Rosenthal
  (2008).
\newblock Selective publication of antidepressant trials and its influence on
  apparent efficacy.
\newblock {\em New England Journal of Medicine\/}~{\em 358\/}(3), 252--260.

\bibitem[\protect\citeauthoryear{Yin and Shi}{Yin and
  Shi}{2019}]{yin2019simulation}
Yin, P. and J.~Q. Shi (2019).
\newblock Simulation-based sensitivity analysis for non-ignorably missing data.
\newblock {\em Statistical methods in medical research\/}~{\em 28\/}(1),
  289--308.

\end{thebibliography}

%%%%%%%%%%%%%%%%%%%%%%%%%%%%%%%%%%%%%%%%%%%%%%%%%%%%%%%%%%%%
\end{document}